# On performance of nano-resonators produced by magnetron sputtering deposition

Mohamed Shaat*

*Engineering and Manufacturing Technologies Department, DACC, New Mexico State University, Las Cruces, NM 88003, USA*

*Mechanical Engineering Department, Zagazig University, Zagazig 44511, Egypt*

## Abstract

Magnetron sputtering is a perfect technique for processing nanomaterials for engineering and medical applications. A material can be processed with specific mechanical properties, microstructure, and surface texture by controlling the parameters of magnetron sputtering. Therefore, studies should be conducted on investigating the processing conditions on the performance of nanomaterials processed by magnetron sputtering.

In this study, effects of the processing force on performance of micro/nano-resonators produced by magnetron sputtering are revealed. The processing force is defined as the ratio of the sputtering power-to-the substrates traveling velocity. By comparing the substrate's traveling velocity to the deposition rate of the sputtered particles, relations are derived for the thickness and surface texture evolutions with the processing force. The coefficients of these relations are experimentally determined for mechanical resonators made of FeNiCr alloy. Then, the variations of the natural frequencies of these resonators with the processing force of magnetron sputtering and the deposition rate of the sputtered particles are depicted. It is demonstrated that special considerations should be given for the effects of the processing conditions when design mechanical resonators produced by magnetron sputtering.

## 1. Introduction

Today's engineering and medical devices incorporate micro/nano-resonators which reflect changes in their properties (*e.g.* frequencies) as a response for physical quantities. For example, nano-resonators are being used for measuring masses of nano-objects (*e.g.* atoms) by measuring the corresponding frequency changes [1-11]. Many factors may affect the accuracy of these resonators in detecting physical quantities [9-11]. Being with ultra-high quality factors is a key factor to easily determine their responses [11-15]. Moreover, it was demonstrated that the microstructure, size, and surface texture (*i.e.* surface roughness, waviness, and altered layer) of these resonators influence their performance in different applications [4, 9, 10, 15]. The operating principle of nanodevices depends on models to transfer a resonator's property change into a value of the measured physical quantity. Thus, to guarantee accurate performance, measures that capture the dependency of the resonators' mechanics on their microstructures, surface textures, and sizes should be introduced in the context of these models.

-------------------------------------------------

∗Corresponding author: *E-mail address:* shaat@nmsu.edu; shaatscience@yahoo.com (M. Shaat).
Tel.: +15756215929

Plasma Vapor Deposition (PVD) is an effective process of producing ultra-thin films for medical and engineering applications. In the context of this process, a positively charged plasma is accelerated by an electrical field generated between a cathode target and an anode substrate. When plasma hits the target, atoms are ejected from the target material to condense on the substrate material. Magnetron Sputtering Deposition is a PVD process in which magnets are used to control the transport of ions between the two electrodes. The characteristics of the growing film strongly depend on the transport of the ions, the transferred momentum, and the deposited energy on the substrate [16-20]. Thus, the characteristics of the film depend on the deposition rate of the sputtered flux [17]. Different models were proposed to estimate the deposition rate of ions in magnetron sputtering (*e.g.* [16-20]). These models help in understanding the transport phenomena of the ions and the characteristics of the growing film. According to these models, the deposition rate depends on the discharge (sputtering) power, the distance between the electrodes, and the processing pressure.

There is a growing interest to investigate the dependency of the material properties on the material processing using magnetron sputtering. Depending on the processing conditions, materials can be obtained with specific mechanical, magnetic and optical properties, microstructures, and surface textures. For example, Gan et al. [21] investigated the structural properties of Cu2O-epitaxial films, Sun et al. [22] determined the optical and electrical performance of thermochromic V2O3 thin films, Cao and Zhou [23,24] determined the magnetic properties of CoZrNb and FeNiCr films, Mirzaee et al. [25] examined the surface textures of ZnO films, Rode et al. [26] studied effects of deposition rate on surface roughness of Al films, Zenkin et al. [27] investigated the thickness dependence of wettability and surface properties of HfO2 thin films, Han et al. [28] explored the composition and structure of $TiH_xHe$ films, Gudla et al. [29] investigated the microstructure evolution of AlZr and AlZrSi coatings during heat treatment processes, Kobata and Miura [30] examined the mechanical and thermal properties of TiCuZrNiHfSi thin films, Pshyk et al. [31] analyzed the structural, morphological and tribo-mechanical properties of AlNTiB2TiSi2 coatings, and Mazur et al. [32] investigated structural, optical, and micro-mechanical properties of (NdyTi1−y)Ox thin films produced by magnetron sputtering. Moreover, effects of the magnetron sputtering processing on the surface roughness of ultra-thin films are investigated in different studies [33-37].

Although studies have been conducted on the characterization of materials processed by magnetron sputtering, the literature lacks for investigations that relate the material performance and mechanics to the magnetron sputtering processing. Therefore, in this study, effects of the processing conditions on the performance of micro/nano-resonators produced by magnetron sputtering are investigated. The frequency characteristics of micro/nano-resonators are related to the target material type, the deposition rate, sputtering power, and the target-substrate relative velocity. It is demonstrated that micro/nano-resonators can be obtained with different thicknesses, surface textures, and frequency characteristics depending on the processing conditions.

## 2. Effects of magnetron sputtering deposition on performance of nano-resonators

A hybrid experimental-analytical mechanics is proposed to relate the surface textures and frequencies of micro/nano-resonators to the parameters of the magnetron sputtering. The configuration of the considered magnetron sputtering is shown in Figure 1. First, the equation of motion of mechanical resonators is derived incorporating measures for the surface texture. Second, the ratio of the sputtering power, $P$, to the substrate's traveling velocity, $v$, is introduced as a processing force. The substrate's traveling velocity is defined as the velocity along the perpendicular direction of the ion transport from the target to the substrate. Then, by comparing the processing force ($P/v$) with the deposition rate, the thickness and surface textures of resonators are related to the magnetron sputtering parameters. Third, by solving the eigenvalue problem of the derived equation of motion, the frequencies of the resonators are obtained as functions of the processing force and the deposition rate. This proposed approach is applied to micro/nano-resonators made of FeNiCr-alloy produced by magnetron sputtering at different processing conditions. Using the experiment, the processing force ($P/v$) along with the deposition rate are determined. In addition, parameters of the surface texture are determined by analyzing the surface of the processed resonators using Atomic Force Microscopy (AFM).

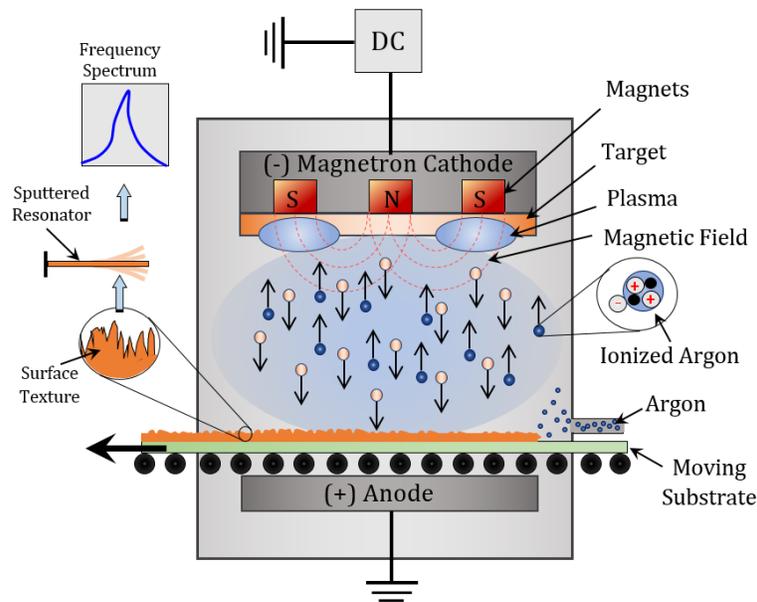

FIG. 1: Configuration of a DC-powered magnetron sputtering. In this configuration, the substrate moves with a uniform velocity, $v$. The processing force is defined as the ratio of the sputtering power to the substrate's traveling velocity ($P/v$).

In this study, a mechanical resonator is modeled as an Euler-Bernoulli beam. The equation of motion under free vibration is derived in the following form [15]:

$$D\frac{\partial^4 w(x,t)}{\partial x^4} + K\frac{\partial^3 w(x,t)}{\partial x^3} + B\frac{\partial^2 w(x,t)}{\partial x^2} + I_0\frac{\partial^2 w(x,t)}{\partial t^2} = 0 \tag{1}$$

where $w$ the resonator's deflection. The stiffnesses and the inertia term in equation (1) are determined as follows:

$$D = E\left(\frac{bh^3}{12} + \frac{2b}{3}\left(\left(R_a + \frac{h}{2}\right) - \frac{h^3}{8}\right)\right)$$

$$K = 4Eb\left(R_a + \frac{h}{2}\right)^2 R_s \tag{2}$$

$$B = 4Eb\left(R_a + \frac{h}{2}\right) R_s^2$$

$$I_0 = \rho b(h + 2R_a)$$

where $E$ is the elastic modulus, $\rho$ is the mass density, $b$ is the width, and $h$ is the thickness of the resonator.

To account for the effects of the surface roughness on the mechanics of the resonator, the stiffnesses, $D$, $B$ and $K$, are introduced depending on the surface average roughness, $R_a$, and the average slope of the surface texture, $R_s$. Equation (1) can be reduced to the one of the classical model that assumes super-polished surfaces by setting $R_a = 0$ and $R_s = 0$.

The thickness of the resonator, $h$, can be related to the deposition rate, $r$, and the traveling velocity of the substrate, $v$, as follows:

$$h = \frac{rd_a}{v} \tag{3}$$

where $d_a$ is the diameter of the deposited particle (*i.e.* if $v = r$, a single-atomic layer is formed at the substrate).

By introducing the ratio of the sputtering power, $P$, to the substrate's traveling velocity, $v$, as the processing force needed to produce a film with a specific thickness, $h$, equation (3) can be rewritten in the form:

$$h = (r_0 d_a)(P/v) \qquad (4)$$

where $r_0$ denotes the deposition rate per a unit sputtering power.

Many factors - including deposition rate, substrate velocity, and machine conditions/vibration - affect the formed surface texture of a sputtered film. The difference between the deposition rate and the substrate velocity is a crucial factor that determines the features of the sputtered surface (*i.e.* $R_a$ and $R_S \propto (r - v)/r$). Therefore, to a great extent, it is acceptable to relate the parameters of the surface roughness, $R_a$ and $R_S$, to the processing force as follows:

$$\begin{aligned} R_a &= \frac{\alpha_a}{(r_0 d_a)} - \frac{\beta_a}{(r_0 d_a)(P/v)} \\ R_s &= \frac{\alpha_s}{(r_0 d_a)} - \frac{\beta_s}{(r_0 d_a)(P/v)} \end{aligned} \qquad (5)$$

where $\alpha_a, \alpha_s, \beta_a$, and $\beta_s$ are coefficients which can be determined using the experiment.

## 3. Application to FeNiCr-resonators produced by magnetron sputtering

In the following, the proposed approach is applied to determine the dependency of the natural frequencies of resonators made of FeNiCr-alloy by magnetron sputtering on the processing force and the deposition rate. To this end, the experiment presented in [24] is reconsidered. In the context of this experiment, a target alloy containing (atomic%) 54 Fe, 38 Ni, and 8 Cr and a polyethylene terephthalate substrate were used at the cathode and anode, respectively. Argon-sputtering gas at 0.6 Pa pressure was supplied where the base pressure was maintained below $8.6 \times 10^{-4}$ Pa. By using different intensities of the sputtering DC-power at a constant velocity of the substrate of $0.5 \, m/min$, films with different thicknesses were obtained. The data of this experiment are presented in Figure 2. Figure 2 shows a linear increase in the thickness of FeNiCr-film with an increase in the processing force $(P/v)$. By fitting these data using equation (4), the deposition rate per a unit sputtering power, $r_0$, of FeNiCr particles is determined by $(d_a r_0 = 458 \, nm/MN)$.

By scanning $1\mu m \times 1\mu m$ area of the surface of the obtained films, the AFM images of the surface textures shown in Figure 3 were obtained. The AFM images are, then, analyzed using *Gwyddion* software. The average roughness and the average slope of the produced surfaces are determined. The profiles of the surface roughness over selected vertical sections of the surface are also depicted.

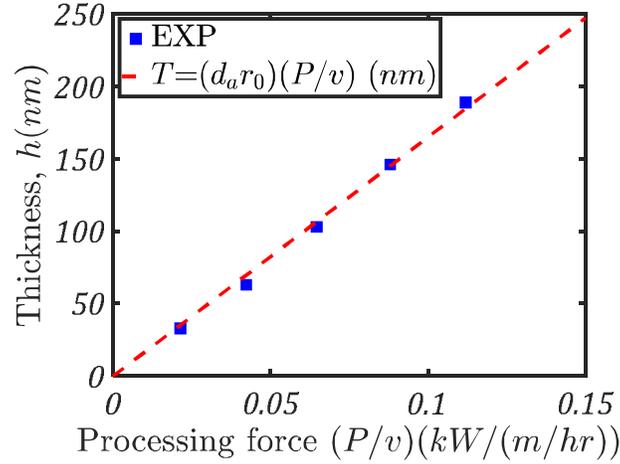

FIG. 2: Thickness-processing force relation for FeNiCr films produced by magnetron sputtering. The squares represent the experimental data (*Adapted from [24]*). The slope of the linear relation is determined as $d_a r_0 = 458 \: nm/MN$.

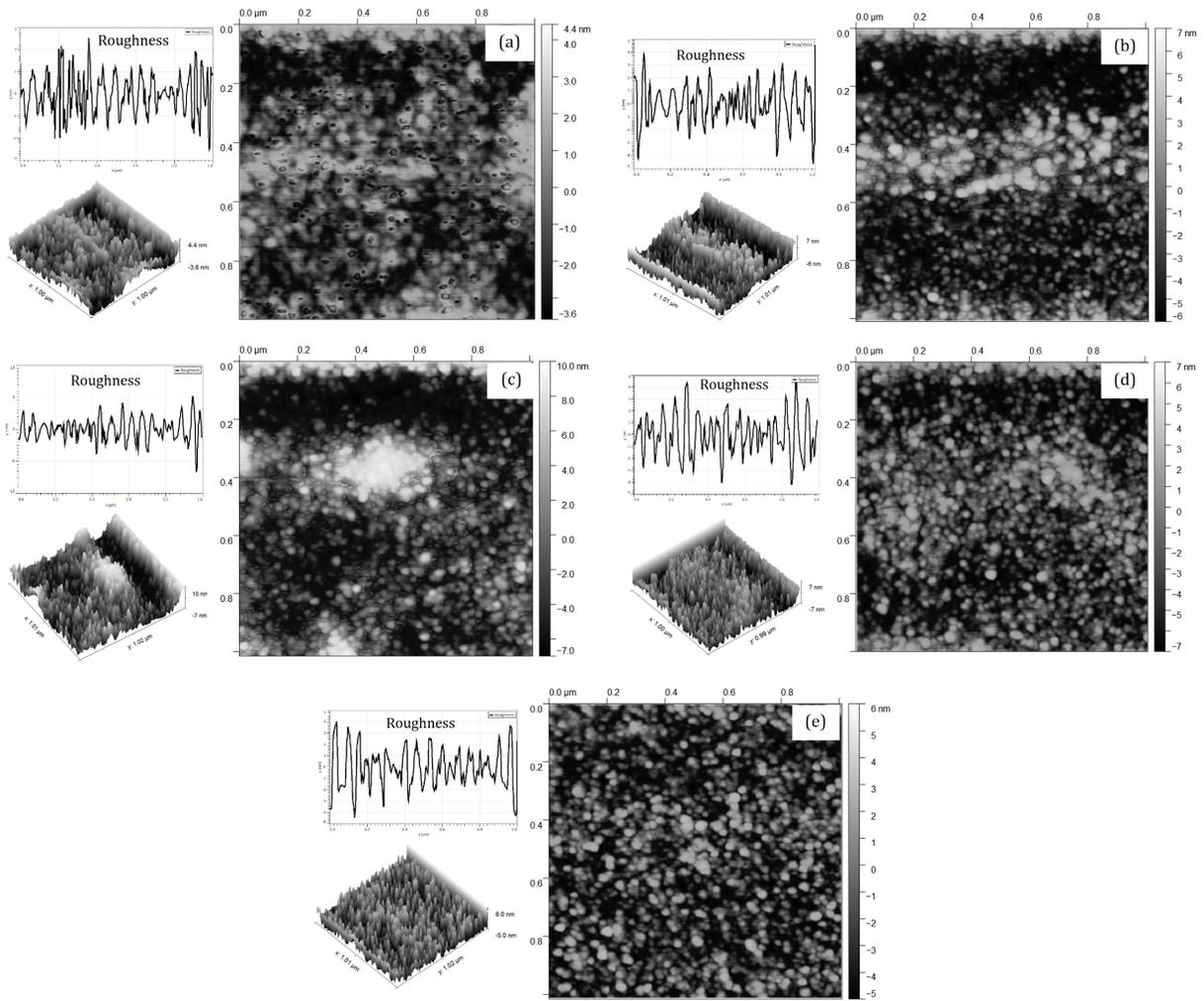

FIG. 3: 2D and 3D AFM images along with profiles of surface roughness of FeNiCr films produced by magnetron sputtering with different processing forces $(P/v)$: (a) $P/v = 76.8 \: kN$, (b) $P/v = 152.4 \: kN$, (c) $P/v = 232.8 \: kN$, (d) $P/v = 316.8 \: kN$, (e) $P/v = 403.2 \: kN$.

As shown in Figure 3, different surface textures are observed when processing FeNiCr by magnetron sputtering with different processing forces. An increase in the range can be observed with an increase in the processing force up to $P/v = 232.8 \ kN$. Moreover, significant changes in the surface textures are observed with an increase in the processing force up to $P/v = 232.8 \ kN$. Utilizing processing forces higher than $P/v = 232.8 \ kN$, similar surfaces with no significant changes are obtained.

Using *Gwyddion* software, the surface average roughness and the average slope of the processed films are determined and plotted against the processing force in Figure 4. By fitting equation (5) with the obtained surface parameters, $\alpha_a = 0.704 \ nm^2/kN$, $\alpha_s = 0.1291 nm/kN$, $\beta_a = 40.9 kN$, and $\beta_s = 42.16 kN$ coefficients are determined. It follows from Figure 4 that the considered relations in equation (5) effectively depict the dependency of the surface parameters on the processing force and the deposition rate.

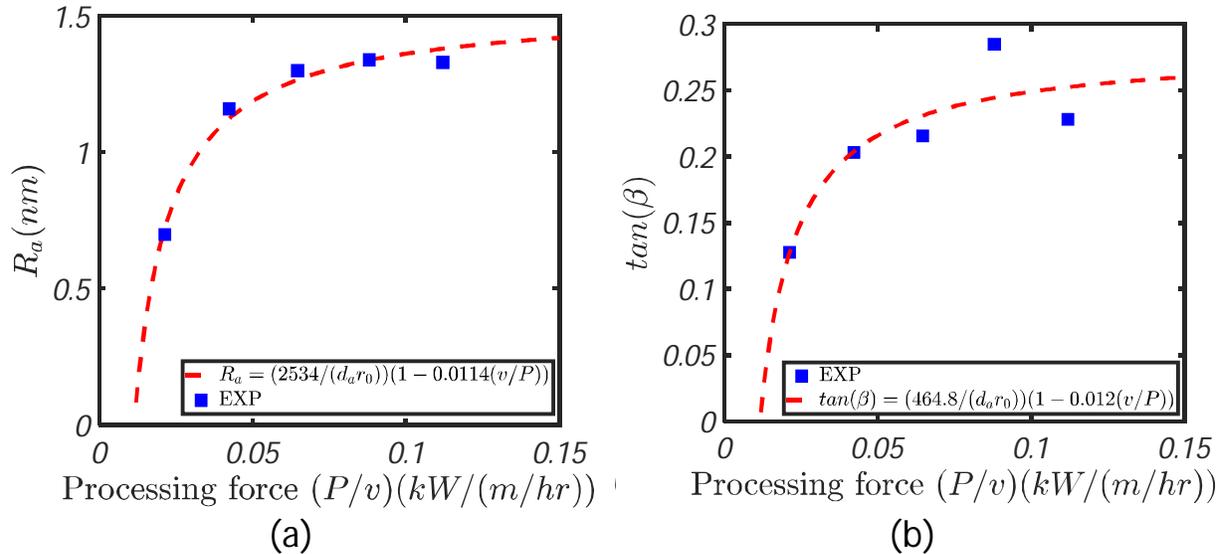

FIG. 4: (a) The average roughness, $R_a$, and (b) the average of the slope of the surface texture, $\tan \beta$, as functions of the driving force ($P/v$). The squares denote the surface parameters as obtained by analyzing the AFM images presented in Figure 3 using *Gwyddion* software. The dash-curves are plotted using equation (5). Note that $R_s = (1 \mu m/L) \tan \beta$, i.e., $L$ denotes the resonator's length.

To report the dependency of the natural frequency of mechanical resonators made of FeNiCr-alloy and processed by magnetron sputtering on the processing force and deposition rate, the eigenvalues of the equation of motion (equation (1)) are obtained. The variations of the nondimensional natural frequencies ($\omega_n = \Omega_n \sqrt{I_0 L^4/D}$, i.e. $\Omega_n$ denote the dimensional natural frequencies) of the first four modes ($n = 1 \rightarrow 4$) as functions of the processing force of the magnetron sputtering are plotted in Figure 5. The figure demonstrates the significant effects of the processing force and processing conditions on the performance of mechanical resonators produced by magnetron sputtering. It is clear that a decrease in the processing force is accompanied with a decrease in the first and second mode-frequencies of a mechanical resonator. For example, resonators produced by magnetron sputtering at low processing forces give very small 1st and 2nd-mode frequencies (may attain zeros). The higher-mode frequencies, on the other hand, are obtained increasing at low processing forces. As the processing force increases, the natural frequencies of the resonator attain the ones of the classical beam with super-polished surfaces. This kind of behavior can be attributed to the fact that surface roughness significantly affect the frequencies of resonators with small sizes. Large sized-resonators, in contrast, are barely affected with their surface textures.

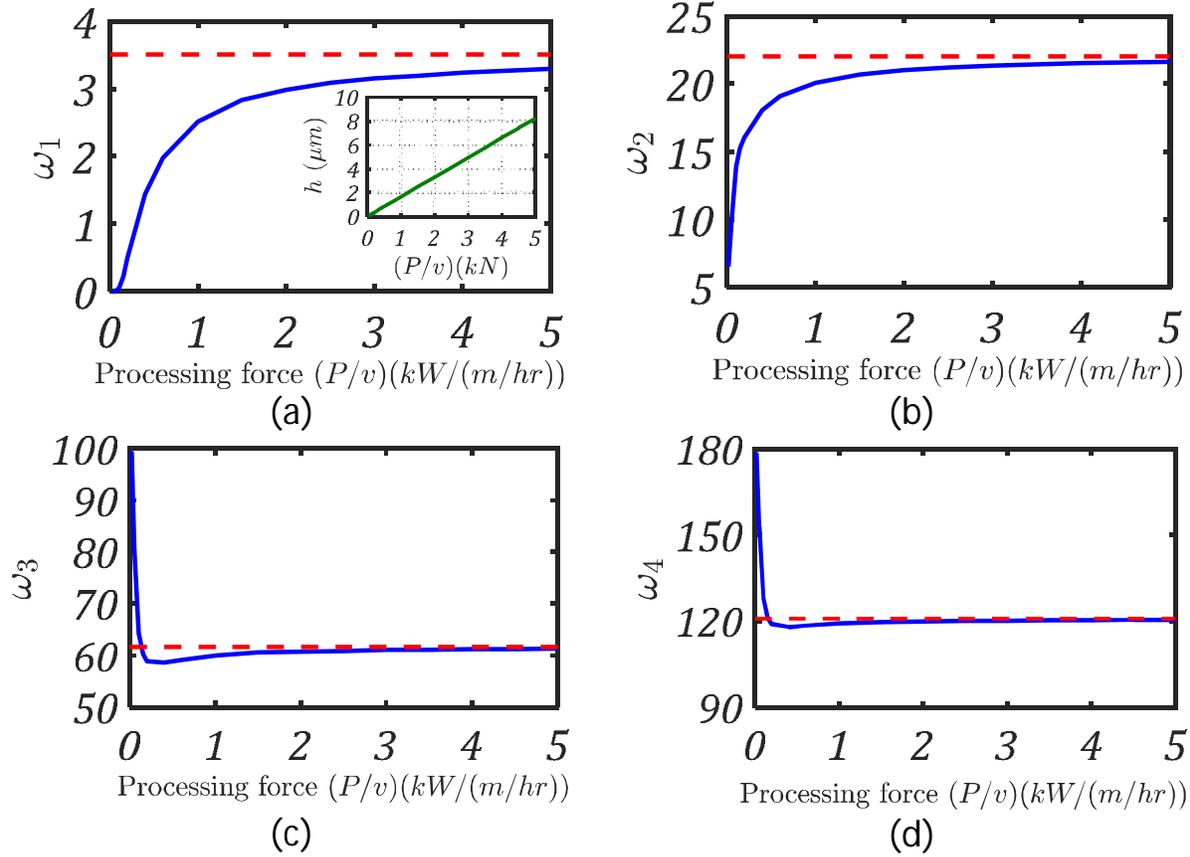

FIG. 5: The first four nondimensional natural frequencies ($\omega_n = \Omega_n \sqrt{I_0 L^4/D}$) of FeNiCr mechanical resonators produced by magnetron sputtering. The figure shows the effects of the processing force ($P/v$) on the frequencies of the resonator. Each of the analyzed FeNiCr resonators is considered with a length of $L = 50h$. The horizontal dash line in each figure represents the nondimensional natural frequency of the corresponding classical beam. The inset in (a) shows the evolution of the resonator's thickness with an increase in the processing force.

## Conclusions

In summary, effects of the processing force on the performance of mechanical micro/nano-resonators produced by magnetron sputtering were revealed. The frequencies of these resonators were obtained as functions of the processing force of magnetron sputtering and the deposition rate of the sputtered flux. It follows from the extracted results that, depending on the processing force of magnetron sputtering, resonators can be fabricated with different thicknesses and surface characteristics. Moreover, magnetron sputtering can produce mechanically-stable resonators by using high processing forces (higher than $3 kW/(m/hr)$). In general, magnetron sputtering should be followed by a surface roughness reduction process (*e.g.* [38, 39]) to enhance the performance of the synthesized resonators.